\documentclass[aps,prl,twocolumn,superscriptaddress,floatfix,letterpaper,nofootinbib]{revtex4-1}

\pdfoutput=1

\usepackage[normalem]{ulem}

\newcommand{\SO}{{\rm SO}}

\newcommand{\U}{{\rm U}}
\newcommand{\SU}{{\rm SU}}

\newcommand{\bea}{\begin{eqnarray}}
\newcommand{\eea}{\end{eqnarray}}

\begin{document}

\begin{titlepage}

\title{
A Solution to the 1+1D Gauged Chiral Fermion Problem 
}

\author{Juven Wang}
\affiliation{School of Natural Sciences, Institute for Advanced Study, Princeton, NJ 08540, USA }
\affiliation{Center of Mathematical Sciences and Applications, Harvard University, MA 02138, USA}
\author{Xiao-Gang Wen}
\affiliation{Department of Physics, Massachusetts Institute of
Technology, Cambridge, MA 02139, USA}

\begin{abstract} 
We show that the 3450 U(1) chiral fermion theory can appear as the low energy
effective field theory of a 1+1D local lattice model, with an on-site U(1)
symmetry and finite-range interactions.  The  on-site U(1) symmetry means that
the U(1) symmetry can be gauged (gaugeable for both background probe and dynamical fields), which leads to a non-perturbative definition
of chiral gauge theory --- a chiral fermion theory coupled to U(1) gauge
theory.  Our construction can be generalized to regularize any
U(1)-anomaly-free 1+1D gauged chiral fermion theory with a zero chiral central
charge (thus no gravitational anomaly) by a lattice, thanks to the recently
proven ``Poincar\'e dual'' equivalence between the U(1) 't Hooft anomaly free condition and the
U(1) symmetric interaction gapping rule, via a bosonization-fermionization
technique.

\end{abstract}


\maketitle

\end{titlepage}

%{\small \setcounter{tocdepth}{1} \tableofcontents }

%\section{Introduction} 

The standard model \cite{G6179,W6764,SW6468,G6267,Z6424,FG7235} is a $\U(1)
\times \SU(2)\times \SU(3)$ gauge theory coupled to fermions that describes all
known elementary particles.
% (except gravitons).  
But until a few years ago, the standard model was only defined perturbatively,
and it is well-known that such a perturbative expansion does not converge.  So
the  standard model is a good way to fit experimental data, but itself was not
a well-defined quantum theory with a finite dimensional Hilbert space for a
finite size system. The reason that the  standard model was not a well-defined
quantum theory is because the left-hand and right-hand fermions in the standard
model carry different $\U(1)\times \SU(2)$ representations.  Such kinds of
fermions are known as chiral fermions \cite{Lee:1956qn}. To define a gauged
chiral fermion theory beyond perturbation is a long-standing problem.
% partly due to the fermion-doubling problem \cite{Nielsen1981}.

There were many previous attempts for the gauge chiral fermion problem.
Lattice gauge theory approach \cite{K7959} fails since it cannot produce low
energy gauged chiral fermions \cite{Nielsen1981}. The Ginsparg-Wilson (GW) fermion approach
\cite{GW8249} has problems since the chiral symmetry \cite{L9842} is 
realized as a non-on-site symmetry
\cite{CLW1141,CGL1314,W1313,W1301,WW1380,Wang:2017loc} and thus is hard to
gauge.  Domain-wall fermion approach \cite{K9242,S9390} also has problems,
since after coupling to gauge fields, the massless gauge bosons will propagate
in one-higher dimension.  The overlap-fermion approach
\cite{NN9362,NN9474,L9995,N0003,S9947,L0128} is a reformulation of domain-wall
fermion approach and face also some problems in a chiral gauge theory.  

In the lattice gauge theory approach, the fermion interactions (except the
gauge interaction) are ignored.  In the mirror fermion approach proposed in
1986 \cite{EP8679,M9259,BMP0628,GP0776,PS10035896}, one started with a lattice
model containing chiral fermions \emph{and} a chiral conjugated mirror sector.
Then, one includes proper direct interaction or boson mediated Swift-Smit
interactions \cite{S8456,S8631} trying to gap out the mirror sector completely,
without breaking the gauge symmetry and without affecting the normal sector.
One proposed condition to gap out the mirror sector is that there are symmetric
mass terms among mirror fermions and composite mirror fermions to give all the
(composite) mirror fermions a mass \cite{EP8679}. However, such a condition can
be satisfied by U(1) anomalous 1+1D chiral mirror fermions which can never be
fully gapped (see the arXiv version of \Ref{W1301}).  This means the
\cite{EP8679}'s criteria is \emph{not} sufficient enough to produce fully
gapped mirror fermions.  The follow-up work
\cite{GPR9396,L9418,GS9409013,CGP1247} failed to demonstrate that interactions
can gap out the mirror sector without  breaking the symmetry in some mirror
fermion models.  It was argued that ``attempts to decouple lattice fermion
doubles by the method of Swift and Smit cannot succeed\cite{BD9216}'' and many
people gave up the mirror fermion approach.

In 2013, \Ref{W1313} pointed out that quantum anomalies are directly connected
to and realized at the boundary of topological orders \cite{W9039} or symmetry
protected topological orders \cite{GW0931,CLW1141,CGL1314} on a lattice in one
higher dimension. This leads to a classification of anomalies
\cite{W1313,KW1458}. From this point of view, the anomaly free condition is
nothing but the condition for the bulk to be a trivial tensor product state.
This leads to a solution of the gauged chiral fermion problem claimed by
Ref.~\cite{W1301}: \emph{For any chiral gauge theory that is free of ALL
anomalies, there exists a local lattice model of the same dimension whose low
energy effective theory realizes the chiral gauge theory.} Here a \emph{chiral
gauge theory} is defined as a boson/fermion theory coupled to gauge fields in a
\emph{weak} coupling limit.  To actually use the above result, \Ref{W1301}
proposed a sufficient condition: \emph{A gauged chiral fermion theory in
$d$-dimensional space-time with a gauge group $G$ is free of all anomalies if
(1) there exist (possibly gauge symmetry breaking) mass terms that make all the
fermions massive, and (2) $\pi_n(G/G_\text{grnd})=0$ for $0\leq n\leq d+1$,
where $G_\text{grnd}$ is the unbroken gauge symmetry group.} 

The above result implies that the $\SO(10)$ gauged chiral fermion theory in the
$\SO(10)$ grand unification \cite{FM7593} can be regulated by a 3+1D lattice.
This way, we solve a long-standing problem of defining the standard model
non-perturbatively \cite{W1301},  at least for a version of standard model from
an $\SO(10)$ grand unification.  Even certain anomalous chiral gauge theories
can be put on a lattice of the same dimension \cite{W1313}. This new way to gap
out the mirror sector was later referred as ``mass without mass terms'' -- a
new mechanism beyond the Higgs mechanism to generate
mass\cite{WW1380,YBX1451,YX14124784,BZ150504312,AC151109071,AC160606312,DW170604648}.
However, the above claim has not been accepted by many people.

In this work, we will provide a supporting evidence that anomaly free gauged
chiral fermion theories can indeed be regularized on the lattice if we allow
lattice fermions to interact properly.  In particular, we show that the
following 1+1D chiral fermion field theory with $\U(1)$ symmetry 
%\begin{align}
\bea
\label{L3450}
\cL = \sum_{a=1}^4
\psi^\dag_{a} (\ii \prt_t - \ii v_a \prt_x) \psi_{a} 
\eea
%\end{align} 
can be realized by a 1+1D lattice model with on-site $\U(1)$ symmetry in low
energies \cite{WW1380,DW170604648}. Here $a=1,2,3,4$ and
$(v_1,v_2,v_3,v_4)=(1,1,-1,-1)$.  Namely, we have two left-moving Weyl fermions
$\psi_{1}$ and $\psi_{2}$ with charge 3 and 4; and two right-moving Weyl
fermions $\psi_{3}$ and $\psi_{4}$ with charge 5 and 0.  We refer the above
$\U(1)$ chiral fermion theory as the 3450 theory. We note that the gauged 3450
chiral fermion theory does not satisfy the above sufficient condition, but
below we show the mechanism of ``mass without mass terms''  still works.

The 1+1D lattice quantum Hamiltonian model that realized the 3450 chiral
fermion theory is given by
\begin{equation}
\label{Hlatt}
\hat H = \sum_{a,i} (\frac{\ii}{2} v_a \hat c_{a,i}^\dag \hat c_{a,i+1} +h.c.) +\hat H_\text{int}
\end{equation}
where $\hat H_\text{int}$ describes the short range interaction whose form will
be given later.  Here the lattice spacing is chosen to be 1.  The $\U(1)$
charges of the lattice fermion operators $\hat c_1, \hat c_2, \hat c_3, \hat
c_4$ are given by $(q_1,q_2,q_3,q_4)=(3,4,5,0)$ and the lattice model has an
on-site $\U(1)$ symmetry: $\hat c_{a,i} \to \ee^{\ii q_a \th} \hat c_{a,i}$.
(An on-site symmetry is a special global symmetry that is a tensor product of
symmetry transformations on each site.  Global symmetries on the lattice with
't Hooft anomalies are necessarily non-on-site
\cite{CLW1141,CGL1314,W1313,W1301,WW1380,Wang:2017loc}.) In this paper, we will
show that, \emph{after choosing the interaction $\hat H_\text{int}$ properly,
the lattice model \eq{Hlatt} will produce the chiral fermion theory \eq{L3450}
at low energies.} This is the key result of the paper.  

In the lattice model \eq{Hlatt}, we do \emph{not} use GW fermion at all, thus
we do \emph{not} encounter the difficulty of gauging GW's non-on-site symmetry
\cite{WW1380,Wang:2017loc}.  Our model \eq{Hlatt}  has an on-site U(1)
symmetry, which  can be easily gauged on a lattice, to obtain a fermionic
lattice model coupled to a U(1) lattice gauge field with a lattice Lagrangian
$L_{\text{latt}}$:
\begin{align}
\label{Llatt-A}
L_{\text{latt}} &= 
\sum_{i,a}  
 c_{a,i}^\dag (\ii \prt_t+ q_a A_{i,0}) c_{a,i}
\nonumber\\
& - \sum_{i,a}  
(\frac{\ii}{2} v_a  c_{a,i}^\dag  c_{a,i+1}\ee^{\ii q_a A_{i,i+1}}  +h.c.) 
- H_\text{int}(A) \nonumber
\end{align}
Now the Lagrangian contains the fermions as Grassmann variables.  The $A_{i,0}$
is a continuous time-component potential, while the $A_{i,i+1}$  is a
spatial-component gauge field on the discretized link.  The $H_\text{int}(A)$
is the gauged version of $H_\text{int}$  based on the standard procedure of
gauging the on-site U(1) symmetry by inserting gauge field $A$ on the links ---
which $H_\text{int}$ will be obtained later in \eqn{Hint}.  Such a lattice
model will produce the gauged chiral fermion theory at low energies.  This way,
we show that the gauged chiral fermion theory can be defined non-perturbatively
via a lattice model.

We like to remark that it is well known that to put the gauged chiral fermion
theory on a lattice, the gauge field does not need to be dynamical.  The gauge
field can be fixed background gauge field.  This is the point of view taken by
this paper: The U(1) gauge field $A_\mu$ is a fixed background gauge field, and
we often fix this background to be $A_\mu=0$.  This is why the  U(1) gauge
field is often not explicit in this paper.

When $\hat H_\text{int}=0$, the low energy effective field theory of the above
lattice model of free fermions is given by
\begin{equation}
\label{L3450full}
\cL = \sum_{a=1}^4
\psi^\dag_{a} (\ii \prt_t - \ii v_a \prt_x) \psi_{a} 
+\psi^\dag_{M,a} (\ii \prt_t + \ii v_a \prt_x) \psi_{M,a} 
\end{equation}
where the chiral fermions $\psi_{a}$ correspond to lattice fermions near
crystal momentum $k=\pi$ and the mirror fermions $\psi_{M,a}$ corresponds
lattice fermions near crystal momentum $k=0$.  We like to stress that the above
effective field theory has a momentum cut-off $\La$ that is of the same order
as the inverse lattice spacing (such as $\La =\frac{\pi}{4}$).  Integrating out
lattice fermions beyond the cut-off scale do not change low energy effective
field theory since the lattice fermions are non interacting.

Eqn. \eq{L3450full} is not a chiral fermion theory since it contains both
normal sector (the chiral fermions sector) $\psi_a$ and the mirror sector
$\psi_{M,a}$.  Next we will try to gap out the mirror sector without breaking
the on-site $\U(1)$ symmetry and without affecting the normal sector, by
introducing interactions that affect only the mirror sector.
Let us first describe the required interaction within the effective field theory
\eq{L3450full}.

Using the null-vector condition \cite{H9590,KCW9963,WW1263} from quantum Hall
edge states \cite{H8285,W9038} (for a more general discussion, see
\Ref{KS10080654,KK11045047,L1309}), we can show that, by choosing a proper
interaction within the mirror sector, we can indeed gap out all the mirror
fermions $\psi_{M,a}$, without breaking the $\U(1)$ symmetry.  To see how to gap
out the mirror fermions, we first bosonize them by introducing eight boson
fields $\phi_a$ and $\phi_{M,a}$, $a=1,\cdots,4$:
\begin{align}
 \ee^{\ii \phi_a}&=\psi_a, &
 \ee^{\ii \phi_{M,a}}&=\psi_{M,a} .
\end{align}
Note that the value of the boson fields is only defined modular $2\pi$.
The bosonized low energy effective field theory is described by (see, for
example, \Ref{W9505}.)
\begin{align}
\cL &=
\frac{1}{4\pi} \Big(
K_{ab} \prt_t \phi_a \prt_x \phi_b - \prt_x \phi_a \prt_x \phi_a
\Big)  
\nonumber\\
&
+\frac{1}{4\pi} \Big(
-K_{ab} \prt_t \phi_{M,a} \prt_x \phi_{M,b} - \prt_x \phi_{M,a} \prt_x \phi_{M,a}
\Big)
\end{align}
where $K$ is diagonal with diag$(K)=(1,1,-1,-1)$.
%The on-site $\U(1)$ transformation is realized as
%\begin{align}
% \phi_1 &\to \phi_1 + 3\; \delta \phi, &
% \phi_2 &\to \phi_2 + 4 \; \delta \phi, 
%\nonumber\\
% \phi_3 &\to \phi_3 + 5 \;\delta \phi, &
% \phi_4 &\to \phi_4 ;
%\nonumber\\
% \phi_{M,1} &\to \phi_{M,1} + 3 \; \delta \phi, &
% \phi_{M,2} &\to \phi_{M,2} + 4 \; \delta \phi, 
%\nonumber\\
% \phi_{M,3} &\to \phi_{M,3} + 5 \; \delta \phi, &
% \phi_{M,4} &\to \phi_{M,4} .
%\end{align}
Now we can introduce an interaction
\begin{align}
 \cL_\text{int} &= \pi \Del V_{ab} \rho_{M,a} \rho_{M,b} +
g_1 \psi_{M,1} (\psi_{M,2}^\dag)^2{}_{pt.s} \psi_{M,3} (\psi_{M,4})^2{}_{pt.s}
\nonumber\\
& \ \ \ \ \
+g_2 (\psi_{M,1}^\dag)^3{}_{pt.s} \psi_{M,2} \psi_{M,3} (\psi_{M,4}^\dag)^3{}_{pt.s}
\nonumber\\
&= 
\frac{1}{4\pi} \Del V_{ab} \prt_x \phi_{M,a}\prt_x \phi_{M,b}
\nonumber\\
&\ \ \ \ \
+
g_1 \cos(l_{1,a} \phi_{M,a}) + g_2 \cos(l_{2,a}\phi_{M,a}),
\label{Lint-fermion}
\end{align}
where 
$\rho_{M,a}=\psi_{M,a}^\dag \psi_{M,a} = \frac{1}{2\pi} \prt_x \phi_{M,a}$
is the density of the mirror
fermions, the fermion point splitting ($pt.s$) is defined as higher derivative term: 
 $(\psi_{M,a})^n_{pt.s}
 \equiv \psi_{M,a}   (\prt_x \psi_{M,a})  \dots (\prt_x^{n-1} \psi_{M,a})$, 
and
\begin{align}
 \v l_1 &= (1,-2,1,2), \ \ \ \v l_2 = (-3,1,1,-3) .
\end{align}
After include the interaction $\cL_\text{int}$, the bosonized low energy effective
Lagrangian becomes
\begin{align}
\label{Lfull}
\cL &=
\frac{1}{4\pi} \Big(
-K_{ab} \prt_t \phi_{M,a} \prt_x \phi_{M,b} - V_{ab}\prt_x \phi_{M,a} \prt_x \phi_{M,b}
\Big)
\nonumber\\
& \ \  \ \ \
+g_1 \cos(l_{1,a} \phi_{M,a}) + g_2 \cos(l_{2,a}\phi_{M,a})
\nonumber\\
& \ \  \ \ \
+\frac{1}{4\pi} \Big(
K_{ab} \prt_t \phi_a \prt_x \phi_b - \prt_x \phi_a \prt_x \phi_a
\Big)  
\end{align}
where $V_{ab}=\del_{ab}+\Del V_{ab}$.
We note that $\v l_1,\v l_2$ are chosen such that $\cL_\text{int}$ has the $\U(1)$
symmetry.  $\v l_1,\v l_2$ also satisfy the null-vector condition
\cite{H9590,KCW9963,WW1263,L1309}
\begin{align}
\label{gapC}
 \v l_1^\top K^{-1}\v l_1 =
 \v l_2^\top K^{-1}\v l_2 =
 \v l_1^\top K^{-1}\v l_2 =0,
\end{align}
which makes the term $g_1 \cos(l_{1,a} \phi_{M,a}) + g_2
\cos(l_{2,a}\phi_{M,a})$ being able to gap out the mirror sector.

In the following, we like to discuss how to choose a proper $V_{ab},g_1,g_2$ in
\eqn{Lfull} to gap out the mirror sector \cite{H9590,KCW9963,WW1263,L1309}. To
understand the dynamics of the interacting mirror fermions and why the mirror
fermions can all be gapped out by the interaction $\cL_\text{int}$, we change
the basis for the $\phi_{M,a}$ field, $ \t \phi_a = W_{ab} \phi_{M,b} $, using
the following GL$(4,\Z)$ transformation
\begin{align}
W=
\bpm
  1&  -2&  1&   2\\
  0&  -3&  2&   2\\
  -5&  2&  2&  -5\\
  3&  -4&  1&   5\\
\epm.
\end{align}
Note that the value of the new boson fields $\t \phi_a$ is still only defined
modular $2\pi$.  In terms of $\t\phi_a$, the bosonized theory can be rewritten
as
\begin{align}
\cL&=
\frac{1}{2\pi}  \prt_t \t \phi_1 \prt_x \t \phi_3 + g_1 \cos(\t \phi_1) 
- \frac{1}{4\pi} \t V_{ab} \prt_x \t \phi_a \prt_x \t \phi_b .
\nonumber\\ &\ \ \ 
+\frac{1}{4\pi} (\prt_t \t \phi_4 \prt_x \t \phi_4 
- \prt_t \t \phi_2 \prt_x \t \phi_2 )
+ g_2 \cos(\t \phi_2-\t \phi_4),
\nonumber\\
 & \text{where }\ \ \  \t V  = (W^{-1})^\top V W^{-1}.
 %\t K  = (W^{-1})^\top K W^{-1},  
\end{align}
We can choose $V_{ab}$ such that
$\t V_{ab} = V_0 \del_{ab}$.
Now $(\t \phi_1,\t \phi_3)$
and $(\t \phi_2,\t \phi_4)$ decouple.

To understand the dynamics of $(\t \phi_1,\t \phi_3)$,
we can integrate out $\t \phi_3$ since it is quadratic, and obtain
\begin{align}
\label{SG}
 \cL&=
\frac{1}{4\pi V_0} (\prt_t \t \phi_1)^2 - \frac{V_0}{4\pi } (\prt_x \t \phi_1)^2 + g_1 \cos(\t \phi_1)
\end{align}
This is the standard sine-Gordon theory 
and a well-known gapped phase of Luttinger liquid.  
Since the operator $\cos(\t \phi_1)$,
having a scaling dimension $1/2$, is relevant, $(\t \phi_1,\t \phi_3)$ are
completely gapped when $g_1\neq 0$.  The energy gap scales as $\Del \sim
|g_1|^{2/3}$.

The dynamics of $(\t \phi_2,\t \phi_4)$
is described by
\begin{align}
\label{FF}
\cL&=
\frac{1}{4\pi} (\prt_t \t \phi_4 \prt_x \t \phi_4 
- \prt_t \t \phi_2 \prt_x \t \phi_2 )
+ g_2 \cos(\t \phi_2-\t \phi_4) 
\nonumber\\ &\ \ \ 
- \frac{V_0}{4\pi} (\prt_x \t \phi_2 \prt_x \t \phi_2 +\prt_x \t \phi_4 \prt_x \t \phi_4) ,
\end{align}
which is the bosonized free fermion theory:
\begin{align}
\cL &= 
\t\psi^\dag_{L} (\ii \prt_t - \ii V_0 \prt_x) \t\psi_{L} 
+\t\psi^\dag_{R} (\ii \prt_t + \ii V_0 \prt_x) \t\psi_{R}
\nonumber\\
&\ \ \ \
+ g_2 (\t\psi^\dag_{R} \t\psi_{L} + h.c.).
\end{align}
$(\t \phi_2,\t \phi_4)$ are gapped out by the fermion mass term $g_2
(\t\psi^\dag_{R} \t\psi_{L} + h.c.)$ when $g_2 \neq 0$.  The energy gap scales
as $\Del \sim |g_2|$.  We see that the fermion interaction $\cL_\text{int}$ can
indeed gap out all the massless modes described by $\phi_a$, provided that we
choose $V_{ab}$ properly.

Next, we like to show that the interaction term $\cL_\text{int}$ can be realized
by an interaction $\hat{H}_\text{int}$ on a lattice.  The key is to introduce lattice
fermion operators
\begin{align}
\hat{\t c}_{a,i} = \sum_{j} f(i-j) \hat c_{a,j}
\end{align}
that have a unit overlap with the fermions in the mirror sector and almost no
overlap with fermions in the normal sector.  This can be easily done since the
fermions in the mirror sector carry $k=0$ crystal momentum and the fermions in
the normal sector carry $k=\pi$ crystal momentum.
For example,  we can choose the function $f(i)$
such that its Fourier transformation is given by
\begin{align}
 \t f(k) \equiv \sum_i \ee^{\ii k i} f(i) = \cos^{2n}(\frac{k}{2}).
\end{align}
Such a function satisfies $f(i)=0$ when $|i|>n$, and thus has a finite range.
Using $\hat{\t c}_{a,i}$, we can design the interaction Hamiltonian as
\begin{align}
\label{Hint}
 &\hat H_\text{int} = \sum_i 
\pi \Del V_{ab} \hat{\t \rho}_{a,i} \hat{\t \rho}_{b,i} + 
\\
& 
\sum_i \Big( g_1 \hat{\t c}_{1,i} (\hat{\t c}_{2,i}^\dag)^2 \hat{\t c}_{3,i} \hat{\t c}_{4,i}^2
+g_2 (\hat{\t c}_{1,i}^\dag)^3  \hat{\t c}_{2,i} \hat{\t c}_{3,i} (\hat{\t c}_{4,i}^\dag)^3  
+h.c. \Big)
\nonumber
\end{align}
where $\hat{\t\rho}_{a,i}=\hat{\t c}_{a,i}^\dag \hat{\t c}_{a,i} $ is the
density operator of the mirror fermions, and $\hat{\t c}_{i}^2,\hat{\t
c}_{i}^3$ are defined again by the point splitting $ \hat{\t c}_{i}^2 \equiv
\hat{\t c}_{i} \hat{\t c}_{i+1}$, $\hat{\t c}_{i}^3 \equiv \hat{\t c}_{i-1}
\hat{\t c}_{i}\hat{\t c}_{i+1}$, \etc.  We note that the lattice interaction
$\hat H_\text{int}$ involve almost only the mirror fermions with crystal
momentum $k\sim 0$.  $\hat H_\text{int}$ hardly involve any fermions in the
normal sector with $k\sim \pi$ and  hardly involve any fermions beyond
$k$-cut-off $\La > \frac{\pi}4$ if $n$ is not too small.  So the lattice
fermions  beyond cut-off remain almost non-interacting.  Integrating out those
fermions give us low energy effective field theory \eqn{Lfull}, where the
mirror sector is shown to be fully gapped if we choose $V_{ab},g_1,g_2$
properly.  This way, we show that 3450 chiral fermion theory can be realized at
low energies of a 1+1D lattice model with a on-site $\U(1)$ symmetry.  In other
words, the chiral fermion theory \eq{L3450} can be fully regularized via a
lattice model \eq{Hlatt} as its low energy effective field theory.   

We remark that the gapped phases of the sine-Gordon theory \eq{SG} and
the free fermion theory \eq{FF} can have correlation lengths much bigger than
the cut-off length when $g_1,g_2$ are small. Therefore,
the above field theory analysis of the gapping process is self-consistent. 
Namely, we can use the bosonized field theory to understand the gapping
process of the lattice mirror fermions.

We can show that there is no additional topological ground state degeneracy (GSD) \cite{WW1263} from the gapped mirror sector, thus GSD=1 
(the energy spectrum within a tiny order $O(e^{-L})$ for a system size $L$).
The only low lying modes are from the gapless chiral sector (the dense energy spectra with a small subgap $O(1/L)$).

The numerical calculation of \Ref{CGP1247} for a particular lattice model
with Yukawa interactions fails to realize the 3450 chiral fermion theory.
However, this does not exclude the possibility that a more carefully designed
1+1D lattice model can realize the 3450 chiral fermion theory. In particular,
to realize a chiral fermion theory, the energy scale of interactions should be
comparable with the kinetic term \cite{W1301,WW1380,DW170604648} (instead of
much bigger than the kinetic term chosen in \Ref{CGP1247}).
%\sout{\cred{and compatible with the gapping rule \cite{WW1380}}.}  

Our approach can be generalized to put any anomaly-free 1+1D $\U(1)$ chiral
fermion theory on a lattice. This is because \Ref{WW1380} proves non-perturbatively
that the $\U(1)$-anomaly free condition with zero chiral central charge $c_L-c_R=0$ (thus no gravitational anomaly)
is equivalent to the $\U(1)$ symmetric interaction gapping rule.
\Ref{WW1380}'s proof is based on the compatibility of anomaly-free condition
and gapping rule \cite{WW1263} under the Narain lattice level quantization
\cite{Narain:1986am} in the context of chiral boson and Chern-Simons
theories. 
For the number of left and right 1+1D Weyl fermions equal $N_L=N_R=N$,
we have constructed the $(\U(1)^N)_{\text{anomaly free}}^{\text{'t Hooft}}$ and 
$(\U(1)^N)_{\text{gapping term}}$ sectors via a short exact sequence in \cite{WW1380}:
$$
{(\U(1)^N)_{\text{anomaly free}}^{\text{'t Hooft}} \to \U(1)^{2N } \to (\U(1)^N)_{\text{gapping term}}}.
$$
The $(\U(1)^N)_{\text{anomaly free}}^{\text{'t Hooft}}$ is the maximal torus group which carries the anomaly-free chiral U(1)$^N$ symmetry.
The $(\U(1)^N)_{\text{gapping term}}$ is the symmetry-breaking group as the \emph{Poincar\'e dual}
U(1)$^N$ symmetry within the total group $\U(1)^{2N}$.
The $(\U(1)^N)_{\text{gapping term}}$ 
corresponds to both the maximal and the minimal set of non-perturbative interaction terms ($N$-linear independent cosine terms, as \eqn{Lint-fermion} which we have $N=2$; whose mathematical concept behind is the so-called \emph{Lagrangian subgroup/submanifold} explained in  \cite{WW1380})
to be included in the mirror sector 
in order to gap the mirror fermions.
Therefore, we can always realize a $\U(1)$-anomaly-free  chiral
matter theory with a zero chiral central charge by a truly local 1+1D
interacting lattice model with an on-site $\U(1)$ symmetry (such as
\eqn{Hlatt}/(\ref{Hint})).

Although we describe our approach using a lattice Hamiltonian formalism, the
result can be applied to the lattice Euclidean path integral
formalism (for Monte Carlo simulation), as long as a proper interaction 
%\eq{Lint-fermion} 
\eq{Hint} is chosen \cite{WW1380}. 
In Ref.~\cite{WW1380}, via a rigorous 1+1D bosonization-fermionization method,
we had performed the exact mapping between
the bosonized theory with the sine-Gordon interaction cosine terms, and the
 fermionized theory with the higher-derivative multi-fermion interaction terms,
  in the continuum field theories and in the regularized lattice models.

As a well-defined quantum theory, our lattice model \eq{Hlatt} has a well
defined UV-complete fermion Green's function.  Although we did not compute this
interacting fermion Green's function on the lattice, such a fermion
Green's function becomes the Green's function of non-interacting free chiral fermions with 3450 U(1) chiral symmetry in
\eqn{L3450} at low energies.

JW thanks Edward Witten for remarks, and thanks Jordan Cotler et al.~for
 discussions.  
JW gratefully acknowledges the support of Corning
Glass Works Foundation Fellowship and NSF Grant PHY-1606531.
X.-G Wen is partially supported by NSF grant DMR-1506475 and DMS-1664412.
This work is also supported by NSF Grant PHY-1306313, PHY-0937443, DMS-1308244, DMS-0804454, DMS-1159412 and Center for Mathematical Sciences and Applications at Harvard University.

\bibliography{all,publst1,ref-add} 
\end{document}